\PassOptionsToPackage{unicode}{hyperref}
\PassOptionsToPackage{hyphens}{url}
\PassOptionsToPackage{dvipsnames,svgnames,x11names}{xcolor}
\documentclass[
  11pt,
]{article}
\usepackage{amsmath,amssymb}
\usepackage{iftex}
\ifPDFTeX
  \usepackage[T1]{fontenc}
  \usepackage[utf8]{inputenc}
  \usepackage{textcomp} 
\else 
  \usepackage{unicode-math} 
  \defaultfontfeatures{Scale=MatchLowercase}
  \defaultfontfeatures[\rmfamily]{Ligatures=TeX,Scale=1}
\fi
\usepackage{lmodern}
\ifPDFTeX\else
\fi
\IfFileExists{upquote.sty}{\usepackage{upquote}}{}
\IfFileExists{microtype.sty}{
  \usepackage[]{microtype}
  \UseMicrotypeSet[protrusion]{basicmath} 
}{}
\makeatletter
\@ifundefined{KOMAClassName}{
  \IfFileExists{parskip.sty}{%
    \usepackage{parskip}
  }{
    \setlength{\parindent}{0pt}
    \setlength{\parskip}{6pt plus 2pt minus 1pt}}
}{
  \KOMAoptions{parskip=half}}
\makeatother
\usepackage{xcolor}
\usepackage[margin=2.5cm]{geometry}
\usepackage{longtable,booktabs,array}
\usepackage{tikz}
\usetikzlibrary{patterns}
\usepackage{calc} 
\usepackage{etoolbox}
\makeatletter
\patchcmd\longtable{\par}{\if@noskipsec\mbox{}\fi\par}{}{}
\makeatother
\IfFileExists{footnotehyper.sty}{\usepackage{footnotehyper}}{\usepackage{footnote}}
\makesavenoteenv{longtable}
\providecommand{\tightlist}{%
  \setlength{\itemsep}{0pt}\setlength{\parskip}{0pt}}
\ifLuaTeX
  \usepackage{selnolig}  
\fi
\IfFileExists{bookmark.sty}{\usepackage{bookmark}}{\usepackage{hyperref}}
\IfFileExists{xurl.sty}{\usepackage{xurl}}{} 
\hypersetup{
  pdftitle={Hardware-rooted attestation for AI-agent evidence: composing IETF RATS with action evidence packages},
  pdfauthor={Anton Sokolov; Researcher, Tyche Institute, Tallinn, Estonia --- ORCID 0000-0003-2452-7096},
  colorlinks=true,
  linkcolor={Maroon},
  filecolor={Maroon},
  citecolor={Blue},
  urlcolor={Blue},
  pdfcreator={LaTeX via pandoc}}

\title{Hardware-rooted attestation for AI-agent evidence: composing IETF
RATS with action evidence packages}
\author{Anton Sokolov\\
\small Researcher, Tyche Institute, Tallinn, Estonia\\
\small ORCID 0000-0003-2452-7096}
\date{Preprint. First deposited June 2026; this version July 2026.\\
Also deposited at Zenodo: doi:10.5281/zenodo.20818671 (CC-BY 4.0).\\
\medskip
\small This work has been submitted to the IEEE for possible publication.
Copyright may be transferred without notice, after which this version may no
longer be accessible.}

\begin{document}
\maketitle

\hypertarget{abstract}{%
\subsection{Abstract}\label{abstract}}

An action evidence package (AEP) is a signed, append-only record of what
an AI agent did, who or what authorised the action, and what the outcome
was. It is a software-layer artefact: it tells a verifier the story of
an action as the agent's own runtime reports it. This note argues that
software attestation of this kind is necessary but not sufficient. When
a verifier's question shifts from ``what does the agent claim it did?''
to ``did the specific model version the operator claims to have deployed
actually produce this output, on unmodified hardware?'', the AEP alone
cannot answer. The missing element is a hardware root of trust: an
attestation that the measured boot and runtime state of the platform
match an endorsed reference. The IETF Remote Attestation Procedures
(RATS) architecture (RFC 9334) and Veraison, an open-source RATS
Verifier implementation (Confidential Computing Consortium / Linux
Foundation), supply exactly this. We propose a \emph{composite
attestation}: hardware Evidence appraised under RATS, bound to a
software AEP. We map a small verifier vocabulary (Authorised /
Unauthorised / Indeterminate / Attested / Contested / Expired) onto RATS
appraisal outcomes, and demonstrate feasibility with a small executed
experiment: on a software Trusted Platform Module (TPM; the swtpm
emulator), an \emph{output-binding} protocol
folds the hash of an AEP outcome and a fresh appraiser nonce into an attestation-key-signed quote, with a model-artefact measurement carried in
a platform register. A minimal RATS-Verifier stand-in resolves the three
platform outcomes as designed --- Attested for a good, fresh quote;
Contested when the model measurement is swapped; Expired when a stale
quote is replayed --- and rejects a forged AEP outcome bound to a valid
quote. The result is a feasibility demonstration on emulated hardware,
not a hardware-rooted guarantee.

\hypertarget{the-question-software-attestation-cannot-answer}{%
\subsection{1. The question software attestation cannot
answer}\label{the-question-software-attestation-cannot-answer}}

Work on evidence instrumentation for AI-governance review has
concentrated, reasonably, on the software layer. An AEP captures the
agent's actions as structured, signed records: an action identifier, the
authorising principal (a human approver, a delegated credential, a
policy decision point), the inputs and tool calls, and the resulting
outcome, all chained so that tampering is detectable after the fact.
This is genuinely useful. It lets an auditor reconstruct \emph{what the
agent's runtime says happened} and check that each action carried a
valid authorisation. It is the layer at which most accountability
questions are actually asked.

But there is a class of question the AEP cannot close on its own.
Consider an operator who states under a regulatory obligation that
production inference runs a particular, audited model version on a
particular hardware configuration. A reviewer who later examines the AEP
sees a faithful, signed account of actions --- but every field in that
account is asserted \emph{by the same software stack whose integrity is
in question}. If the runtime was silently swapped for a different model,
or if the host was modified to disable a safety filter, the AEP can be
perfectly well-formed and perfectly signed and still describe a world
that does not correspond to the hardware that ran. The signature proves
the record was produced by a key the runtime holds; it does not prove
what the runtime \emph{was}.

This is not a hypothetical gap. It is the standard limitation of any
self-report: the witness is also the suspect. The AEP answers ``what was
authorised and what was the outcome''; it does not answer ``was the
platform that produced this record the platform the operator claims, in
the state the operator claims''. Closing that second question requires
evidence that originates \emph{below} the software the operator controls
--- evidence rooted in hardware the operator cannot rewrite.

The contribution of this note is the composition, not a new mechanism:
\S\S2--4 locate the AEP against the RATS architecture and specify the
output-binding, \S5 maps a six-term verifier vocabulary onto RATS
appraisal outcomes, \S6 reports a small executed feasibility experiment
on an emulated TPM, and \S7 states scope, limits, and relation to prior
work.

\hypertarget{the-flight-recorder-analogy}{%
\subsection{2. The flight-recorder
analogy}\label{the-flight-recorder-analogy}}

A useful intuition comes from aviation. A flight data recorder is sealed
by its manufacturer, records continuously during operation, and is
opened only later, under supervision, by an independent investigator.
Its value rests on two properties that have nothing to do with the
contents of any single recording: the device is built and sealed by a
party other than the operator, and the conditions under which it is
opened and read are controlled by a party other than the operator. The
recording is trustworthy not because the crew vouches for it but because
the recorder's provenance and custody are independent of the crew.

An AEP, by contrast, is more like a logbook the crew writes themselves.
It can be honest, careful, and tamper-evident --- but its
trustworthiness ultimately traces back to the operator's own runtime. A
hardware-rooted attestation supplies the missing flight-recorder
properties: a measurement produced by a component the operator did not
author (a trusted execution environment or hardware security module),
endorsed by the hardware vendor, and read by an independent appraiser.
The composite we propose pairs the crew's logbook with a
manufacturer-sealed recorder. Neither replaces the other. The logbook
(AEP) carries the semantics --- who authorised what, and to what end.
The recorder (hardware Evidence) carries the assurance that the platform
writing the logbook was the one it claims to be.

\hypertarget{what-rats-contributes}{%
\subsection{3. What RATS contributes}\label{what-rats-contributes}}

The IETF Remote Attestation Procedures working group defines an
architecture for exactly this kind of independent, hardware-rooted
appraisal. The architecture document is RFC 9334 {[}1{]}. Its central
abstraction is a set of roles and a flow of \emph{conceptual messages}
between them:

\begin{itemize}
\tightlist
\item
  An \textbf{Attester} produces \textbf{Evidence} about its own state
  --- for example, a quote over platform measurement registers, signed
  by a key rooted in a hardware component.
\item
  A \textbf{Verifier} appraises that Evidence against \textbf{Reference
  Values} (the expected, ``known-good'' measurements) and
  \textbf{Endorsements} (statements from the hardware vendor about the
  Attester's capabilities and keys), and emits \textbf{Attestation
  Results}.
\item
  A \textbf{Relying Party} consumes the Attestation Results and makes a
  trust decision.
\end{itemize}

The separation that matters here is between the Verifier and the Relying
Party. In a self-report, those collapse into the same software stack.
RATS pulls them apart: the party that decides whether to believe the
measurements (Verifier) is architecturally distinct from the party that
holds the secret being measured (Attester) and from the party that acts
on the verdict (Relying Party). That is the structural property the
flight recorder has and the logbook lacks.

On the underlying hardware mechanism, we state only what we are
confident about and defer the rest. A Trusted Platform Module accumulates
measurements into Platform Configuration Registers (PCRs) during a
measured boot, such that the final register values are a function of the
exact sequence of components loaded; a quote signs those register values
with a key the TPM holds. The Verifier compares the quoted PCR values
against reference values for the configuration the operator claims to
run. The register semantics and command set are defined in the Trusted
Computing Group (TCG) \emph{Trusted Platform Module Library
Specification, Family 2.0}
{[}TCG-TPM2{]}; the firmware-to-PCR measurement procedure and event-log
format are defined in the TCG \emph{PC Client Platform Firmware Profile
Specification, Family 2.0} (v1.06 Rev 52, 4 Dec 2023) {[}TCG-PCCFW{]}.
RATS also defines standard claim formats for conveying Evidence and
Results; the principal token format is the Entity Attestation Token
(EAT), specified in RFC 9711 (Proposed Standard, 2025) {[}3{]}. Beyond
the architecture (RFC 9334) and EAT (RFC 9711), the attestation-results
vocabulary and its serialisation are still being standardised as
Internet-Drafts --- Attestation Results for Secure Interactions (AR4SI)
{[}AR4SI{]} and EAT Attestation Results (EAR) {[}EAR{]} --- and are
cited here as Internet-Drafts, not as RFCs.

Veraison is a relevant open-source Verifier implementation: an
attestation Verifier project developed in the Confidential Computing
Consortium / Linux Foundation ecosystem, that ingests Evidence and
reference/endorsement material and produces Attestation Results {[}2{]}.
(RFC 9334 is an Informational architecture document and names no single
canonical reference implementation; Veraison is one widely-used
open-source Verifier, not ``the'' reference implementation.) We are
confident Veraison exists and plays the Verifier role; we are \emph{not}
confident about specific service names, endpoint paths, or
request/response field names, and we will not invent them. Where the
text reaches a concrete API surface, it names the layer and leaves the
identifier to the project's current repository rather than risking a
stale or invented name.

\hypertarget{the-composite-attestation}{%
\subsection{4. The composite
attestation}\label{the-composite-attestation}}

The proposal is to bind two artefacts that today live in separate
worlds:

\begin{enumerate}
\def\labelenumi{\arabic{enumi}.}
\tightlist
\item
  \textbf{Hardware Evidence} --- a RATS Attester quote over the
  platform's measured state (boot measurements; ideally a measurement or
  identity of the loaded model artefact), appraised by an independent
  Verifier such as Veraison, yielding Attestation Results.
\item
  \textbf{Software AEP} --- the agent's signed action record: action,
  authorising principal, inputs/tool calls, outcome, chained for
  tamper-evidence.
\end{enumerate}

The binding is what gives the composite its force. The AEP, at the point
an action is recorded, should incorporate a reference to a fresh
Attestation Result (or to the Evidence and the appraisal that produced
it), so that a later reviewer can ask not only ``was this action
authorised and what was its outcome?'' but ``and was it produced on a
platform whose state was independently attested at the time?''.
Mechanically, the AEP entry can carry the Attestation Result's
identifier and a hash commitment, and the freshness mechanism (a nonce
from the Relying Party, or a timestamp policy) ties the attestation to
the action's time window rather than to some stale earlier boot. RFC
9334 §10 {[}1{]} defines three such mechanisms --- synchronised-clock
timestamps, nonces, and epoch identifiers --- and the AEP binding can
use a Relying-Party nonce or an epoch/timestamp policy accordingly.

The division of labour is clean:

\begin{itemize}
\tightlist
\item
  The \textbf{AEP} answers \emph{authorisation and outcome} questions.
  It is the semantic layer.
\item
  The \textbf{hardware Evidence + RATS appraisal} answers the
  \emph{platform-state} question. It is the assurance layer.
\item
  The \textbf{composite} lets a single verifier traverse from ``this
  output'' to ``produced by this attested platform, under this
  authorisation, with this outcome'' without trusting the operator's
  word for any link in the chain.
\end{itemize}

Neither layer subsumes the other. A perfect RATS appraisal
says nothing about whether an action was authorised; a perfect AEP says
nothing about whether the platform was genuine. Accountability for AI
agents needs both, and needs them bound.

\hypertarget{a-small-verifier-vocabulary-mapped-onto-rats}{%
\subsection{5. A small verifier vocabulary mapped onto
RATS}\label{a-small-verifier-vocabulary-mapped-onto-rats}}

For the composite to be usable by a non-specialist reviewer, the
appraisal needs to resolve to a small, legible vocabulary rather than a
wall of register values. We propose six terms, grouped by the question
they answer.

\textbf{Authorisation axis (from the AEP):}

\begin{itemize}
\tightlist
\item
  \textbf{Authorised} --- the action carried a valid authorising
  principal under policy at the time it was taken.
\item
  \textbf{Unauthorised} --- the action lacked a valid authorisation, or
  the authorisation was invalid/revoked.
\item
  \textbf{Indeterminate} --- the AEP is incomplete or ambiguous;
  authorisation can be neither confirmed nor refuted.
\end{itemize}

\textbf{Platform axis (from RATS appraisal):}

\begin{itemize}
\tightlist
\item
  \textbf{Attested} --- Evidence appraised successfully against
  reference values and endorsements; the platform state matches the
  operator's claim.
\item
  \textbf{Contested} --- Evidence was produced and appraised, but it
  \emph{fails} to match reference values (e.g.~PCR mismatch indicating a
  modified boot or a different model artefact). This is the high-signal
  case: not absence of evidence, but evidence of a discrepancy.
\item
  \textbf{Expired} --- Evidence is stale relative to the freshness
  policy, or supporting material (endorsement, reference value, signing
  key) has lapsed; the appraisal cannot be trusted as current even if it
  once matched.
\end{itemize}

The mapping tracks two Internet-Drafts rather than published RFCs. AR4SI
{[}AR4SI{]} defines four trustworthiness tiers --- None, Affirming,
Warning, Contraindicated --- serialised as an EAR {[}EAR{]}. Two of the
platform terms correspond directly to those tiers: an \emph{affirming}
appraisal \(\rightarrow\) \textbf{Attested}; a \emph{warning} or
\emph{contraindicated} appraisal that runs but contradicts reference
values \(\rightarrow\) \textbf{Contested}; while the \emph{none} tier,
in which the Verifier asserts nothing, denotes an inconclusive appraisal
rather than a pass or a fail. \textbf{Expired is deliberately not one of
the AR4SI tiers}: it captures a separate, token-level condition ---
evidence stale relative to the freshness policy, or supporting material
(endorsement, reference value, signing key) that has lapsed --- surfaced
by the EAT \texttt{exp} claim and by nonce-based evidence freshness, not
by the trustworthiness vocabulary. Because AR4SI and EAR are still
evolving Internet-Drafts, this correspondence is presented as
provisional and should be validated against a Verifier's actual EAR
output. The authorisation axis is computed entirely from the AEP and
policy and does not depend on RATS.

A composite verdict is then a pair, e.g.~\emph{(Authorised, Attested)}
for an action that was both properly authorised and produced on an
attested platform, or \emph{(Authorised, Contested)} --- the most
interesting failure --- where the action \emph{looks} authorised in the
logbook but the hardware says the platform was not what the operator
claimed. The flight-recorder framing earns its keep here:
\emph{(Authorised, Contested)} is precisely the case where the crew's
logbook reads clean but the sealed recorder disagrees. Figure~\ref{fig:matrix}
lays out the full verdict space and marks the cells the experiment of \S6
exercises.

\begin{figure}[t!]
\centering
\begin{tikzpicture}[x=3.4cm,y=1.5cm]
  \foreach \i in {0,1,2,3} { \draw[black] (\i,0) -- (\i,-3); }
  \foreach \j in {0,1,2,3} { \draw[black] (0,-\j) -- (3,-\j); }
  \node[font=\bfseries\small] at (0.5, 0.35) {Attested};
  \node[font=\bfseries\small] at (1.5, 0.35) {Contested};
  \node[font=\bfseries\small] at (2.5, 0.35) {Expired};
  \node[font=\small\itshape] at (1.5, 0.85) {Platform axis (RATS appraisal of the quote)};
  \node[font=\bfseries\small, anchor=east] at (-0.12,-0.5) {Authorised};
  \node[font=\bfseries\small, anchor=east] at (-0.12,-1.5) {Unauthorised};
  \node[font=\bfseries\small, anchor=east] at (-0.12,-2.5) {Indeterminate};
  \node[font=\small\itshape, rotate=90, anchor=south] at (-0.95,-1.5) {Authorisation axis (AEP + policy)};
  \fill[pattern=north east lines] (1.02,-0.98) rectangle (1.98,-0.02);
  \draw[line width=1.6pt] (0,0) rectangle (1,-1);
  \draw[line width=1.6pt] (1,0) rectangle (2,-1);
  \draw[line width=1.6pt] (2,0) rectangle (3,-1);
  \node[font=\small, fill=white, inner sep=2pt] at (0.5,-0.5) {Case A};
  \node[font=\small, fill=white, inner sep=2pt] at (1.5,-0.5) {Case B};
  \node[font=\small, fill=white, inner sep=2pt] at (2.5,-0.5) {Case C};
  \foreach \cx in {0.5,1.5,2.5} { \foreach \cy in {-1.5,-2.5} {
    \node[font=\footnotesize\itshape] at (\cx,\cy) {valid; not exercised};
  } }
  \node[font=\footnotesize, anchor=north, align=center] at (1.5,-3.25)
    {thick border = exercised in \S6 \quad hatching = flagship failure\\
     \textit{valid; not exercised} = reachable verdict, outside this run};
\end{tikzpicture}
\caption{The composite verdict space: the authorisation axis (computed
from the AEP and policy alone) crossed with the platform axis (the RATS
appraisal of the hardware quote). All nine pairs are reachable composite
verdicts. The \S6 feasibility run holds the authorisation axis fixed ---
its synthetic AEP is authorised by construction --- because that axis is
computed from the AEP and policy alone and needs no hardware to
exercise; the run varies only the platform axis, giving the three
exercised cells (Table~\ref{tab:verdicts}). Hatching marks the flagship
failure \emph{(Authorised, Contested)} --- the logbook reads clean, the
sealed recorder disagrees. The binding check sits below the matrix: a
forged AEP outcome presented under a valid quote is rejected before any
verdict is assigned.}\label{fig:matrix}
\end{figure}

\hypertarget{a-feasibility-experiment-executed}{%
\subsection{6. A feasibility experiment
(executed)}\label{a-feasibility-experiment-executed}}

We ran a small, reproducible feasibility experiment to make the binding
concrete. It is deliberately modest: a single developer host, a
\emph{software} TPM, and a minimal RATS-Verifier stand-in. Its purpose
is to show that the output-binding protocol and the three-way platform
verdict are realisable end to end --- not to make any hardware-rooted
security claim.

\textbf{Setup.} A software TPM (swtpm 0.7.3, libtpms-based {[}swtpm{]})
stands in for a hardware TPM on a single developer workstation;
tpm2-tools 5.6 drives it. A simulated measured boot extends platform
configuration registers (PCRs) 0--3, with \textbf{PCR 3 carrying a
stand-in measurement for the model artefact} --- a tagged digest
extended into the register rather than a hashed model binary --- so that
the model version is, by design, part of the attested state. A synthetic
AEP is built, and the proposed \emph{output-binding} is applied: the
SHA-256 digest of the AEP outcome together with a fresh appraiser nonce
is folded into the qualifying data of an attestation-key-signed TPM
quote. The appraiser is an explicit \emph{stand-in} for a full Verifier
such as Veraison: it performs the signature and
freshness/qualifying-data check (\texttt{tpm2\_checkquote}) and a
reference-value comparison of the quoted PCRs against a recorded good
state. It does not implement endorsement chains, Concise Reference Integrity
Manifest (CoRIM) reference values,
or the AR4SI/EAR result serialisation.

\textbf{Results.} The AEP under test has digest
\texttt{b80c48d9\ldots{}e59d89} (the full SHA-256 value is recorded in the
run's \texttt{results.json}; see Data and code availability). The four
appraisal cases resolve as designed (Table~\ref{tab:verdicts}):

\begin{longtable}[]{@{}
  >{\raggedright\arraybackslash}p{(\columnwidth - 4\tabcolsep) * \real{0.12}}
  >{\raggedright\arraybackslash}p{(\columnwidth - 4\tabcolsep) * \real{0.44}}
  >{\raggedright\arraybackslash}p{(\columnwidth - 4\tabcolsep) * \real{0.44}}@{}}
\caption{The four appraisal cases of the feasibility run and their
verdicts. Cases A--C exercise the three platform-axis terms of \S5;
the binding case shows that a forged AEP outcome cannot ride on a valid
quote.}\label{tab:verdicts}\\
\toprule\noalign{}
Case & Condition & Verdict \\
\midrule\noalign{}
\endhead
\bottomrule\noalign{}
\endlastfoot
A & good state, fresh nonce & \textbf{Attested} --- signature valid,
PCRs match the reference \\
B & PCR 3 perturbed --- a different digest extended in, standing in for
a swapped model artefact & \textbf{Contested} --- signature valid, but
PCRs diverge from the reference \\
C & run-A quote replayed under a new nonce challenge & \textbf{Expired}
--- the nonce-freshness check fails \\
binding & a forged AEP outcome presented under run-A's quote &
\textbf{rejected} --- \texttt{tpm2\_checkquote} fails; the outcome
cannot be swapped without invalidating the quote \\
\end{longtable}

Cases A--C exercise each of the three platform-axis terms from §5 and
confirm they are reachable and distinguished. Case B is the load-bearing
one: a perturbed model-artefact measurement surfaces as
\textbf{Contested} rather than silently passing, which is exactly the
\emph{(Authorised, Contested)} failure the composite exists to make
visible. The binding case shows the output-binding does its job --- an
attacker cannot retain a valid quote while substituting a different AEP
outcome.

\textbf{Scope and threats to validity.} Five caveats bound these claims,
and we state them plainly. First, this is \textbf{software-TPM
emulation}: it demonstrates the protocol and the binding, not a genuine
hardware root of trust; conclusions about real, unmodified-hardware
guarantees cannot be drawn from an emulated Attester. Second, the
appraiser is a \textbf{minimal stand-in} for a real RATS Verifier --- it
performs the checks a Verifier performs (signature, freshness,
reference-value comparison) but not endorsement appraisal, CoRIM
ingestion, or AR4SI/EAR result production; the exact Veraison
service and interface names are deferred to the project's current
repository, to be fixed when a real appraisal is wired up. Third, the
AEP is \textbf{synthetic} and exercises the binding mechanics, not real
agent semantics. Fourth, the §5 vocabulary-to-result-schema mapping
remains provisional until validated against a real Verifier's EAR
output. Fifth, the run exercises one sub-case per term: Case C fails
only the nonce-freshness check on a replayed quote, while the EAT
\texttt{exp} and lapsed-endorsement paths that §5 also folds into
Expired are not exercised; and Case B perturbs a stand-in digest in PCR
3 rather than re-measuring an actual model binary. A hardware-rooted
study --- a real TPM/TEE, a full Veraison appraisal, and a genuinely
measured model artefact --- is the next step.

\textbf{Determinism and reproducibility.} A second-order question is
what the binding establishes when inference is not deterministic. The
output-binding here commits the AEP \emph{outcome} digest, not the
model's token-level generation: it makes the pairing of a particular
outcome with a particular attested platform state tamper-evident within
the freshness window, and that provenance holds whether the model
decoded greedily or sampled stochastically. What stochastic decoding
removes is \emph{independent re-derivation}. A third party cannot re-run
the model on the recorded input and reproduce the same output, so the
binding shows that this outcome was bound to this attested state --- not
that it is the unique output of this model on that input. Recovering the
stronger property would mean measuring the decode parameters and the
sampler's PRNG state at the start of generation into the attested state
alongside the model artefact; the seed alone suffices only when the
generator is reset under a fixed algorithm. Even then, re-derivation
holds only over a pinned numerical pipeline, because parallel and
accelerator reductions make floating-point results non-associative. The
experiment above used a fixed, deterministic outcome and does not
exercise this case.

\hypertarget{scope-limits-and-relation-to-prior-work}{%
\subsection{7. Scope, limits, and relation to prior
work}\label{scope-limits-and-relation-to-prior-work}}

This note deliberately stays at the architecture-and-design level. It is
a companion to the author's evidence-instrumentation work {[}4{]}, which
treats the software AEP as a first-class governance artefact; the contribution
here is to locate that artefact correctly in the trust stack --- as the
necessary software layer --- and to specify what a hardware layer
beneath it would add and how the two compose.

The mechanism itself is deliberately not new, and the claim of this
note should be read against its lineage. Folding application data into
a quote's qualifying data is the standard output-binding pattern:
Intel SGX carries user data in the report's \texttt{report\_data} field
for exactly this purpose {[}SGX{]}, and continuous appraisal of
attestation-key-signed TPM quotes against reference values is the
operating principle of Keylime {[}Keylime{]}, the closest existing
system --- what Keylime does for node integrity, the composite here
does for a single governance record. The engineering path for the model
measurement this note leaves as a caveat also exists: Linux IMA descends
from the TCG-based integrity-measurement architecture of Sailer et
al.\ {[}IMA{]}. And a complementary research line answers the same
operator-honesty question by different means: verifiable inference
proves properties of the \emph{computation} --- zk-SNARK proofs of
model execution {[}zkML{]}, or confidential-computing GPUs with native
device attestation {[}HCC{]} --- where composite attestation proves
properties of the \emph{platform} and binds the governance record to
them. The two are compatible and fail differently. What is new here is
the composition alone: an application-layer action record bound into
RATS Evidence and resolved, with the platform verdict, into one
two-axis vocabulary.

Three honest limitations. First, hardware attestation moves the trust
question rather than dissolving it: one must now trust the hardware
vendor's endorsements and the Verifier's independence, and the
supply-chain and key-management assumptions behind those are real and
non-trivial. Second, attesting a \emph{model version} specifically ---
as opposed to a boot configuration --- requires that the model artefact
be measured into the attested state, which is an engineering commitment
not all deployments make today. Third, the RATS/TCG specifics are now
cited at their current versions (RFC 9334; RFC 9711; AR4SI and EAR as
Internet-Drafts; the TCG TPM 2.0 Library and PC Client Firmware Profile;
swtpm), with two items deliberately left open --- the exact Veraison
service/interface names (to be fixed against the current repository
when a real appraisal is wired up) and the §5
result-schema-to-vocabulary mapping (to be validated against a
Verifier's EAR output) --- because getting an RFC
number or an API field wrong would undermine a note whose whole point is
verifiable provenance.

The framing of eIDAS, PKI, and trust services in this note is the
author's research expertise and observation, not a service offered.

\hypertarget{conclusion}{%
\subsection{8. Conclusion}\label{conclusion}}

An action evidence package tells a verifier what an agent did and under
whose authority, and tells it from inside the very stack whose
integrity may be in doubt. That is enough for many accountability
questions and not enough for the sharpest one: \emph{was this the
platform, in the state, that the operator claims?} IETF RATS (RFC 9334)
and Veraison answer that question with an independently appraised,
hardware-rooted Evidence claim. Composing the two (software AEP for
authorisation and outcome, hardware Evidence for platform state, bound
at record time and resolved into a six-term vocabulary) gives a
reviewer a path from an output back to an attested, authorised origin
without taking the operator's word for any step. The experiment reported
here is small and software-TPM-emulated by design; it demonstrates the
binding and the three-way platform verdict end to end (Attested /
Contested / Expired, with a forged outcome rejected), and its remaining
limits --- a real hardware root, a full Veraison appraisal, a genuinely
measured model --- define the next study rather than the claim made
here.

\hypertarget{declarations}{%
\subsection{Declarations}\label{declarations}}

\textbf{Generative-AI assistance.} The author used Anthropic's Claude
for drafting assistance, literature triage, and editorial review; all
claims, results, design decisions, and references are the author's own
and verified by the author, who takes full responsibility. Reported per
COPE and ICMJE recommendations.

This work received no external funding. The author declares no
competing interests. The work involved no human or animal subjects.

\textbf{Data and code availability.} The run script
(\texttt{run\_pipeline.sh}), the run log, the decoded results
(\texttt{results.json}, carrying the full SHA-256 digest reported in
\S6), the AEP fixture (\texttt{aep.json}), and the reference PCR values
underlying \S6 are retained by the author and available on request; a
public deposit of the run package is planned alongside the next,
hardware-rooted study. The tools used are all open source (swtpm,
tpm2-tools; see References).

\hypertarget{references}{%
\subsection{References}\label{references}}

\emph{RFC 9334 and RFC 9711 are confirmed against the IETF datatracker.
AR4SI and EAR are cited as live Internet-Drafts; versions as of June
2026.}

{[}1{]} H. Birkholz, D. Thaler, M. Richardson, N. Smith, W. Pan,
``Remote ATtestation procedureS (RATS) Architecture,'' RFC 9334, IETF,
January 2023 (Informational). Freshness mechanisms: §10.

{[}2{]} Project Veraison (VERificAtIon of atteStatiON), open-source RATS
attestation-verification components, originated at Arm and donated to
the Confidential Computing Consortium (Linux Foundation).
https://github.com/veraison

{[}3{]} L. Lundblade, G. Mandyam, J. O'Donoghue, C. Wallace, ``The
Entity Attestation Token (EAT),'' RFC 9711, IETF, 2025 (Proposed
Standard).

{[}4{]} A. Sokolov, ``Evidence instrumentation for AI-governance review:
underlying data and extended materials for three public-source
practitioner use cases,'' Zenodo, 2026 (dataset, CC BY 4.0).
doi:10.5281/zenodo.20488643.

{[}Keylime{]} N. Schear, P. T. Cable II, T. M. Moyer, B. Richard, R.
Rudd, ``Bootstrapping and Maintaining Trust in the Cloud,'' Proc.\ 32nd
Annual Conference on Computer Security Applications (ACSAC), 2016, pp.
65--77. doi:10.1145/2991079.2991104.

{[}IMA{]} R. Sailer, X. Zhang, T. Jaeger, L. van Doorn, ``Design and
Implementation of a TCG-based Integrity Measurement Architecture,''
Proc.\ 13th USENIX Security Symposium, 2004.

{[}SGX{]} V. Costan, S. Devadas, ``Intel SGX Explained,'' IACR
Cryptology ePrint Archive, Report 2016/086, 2016.

{[}zkML{]} D. Kang, T. Hashimoto, I. Stoica, Y. Sun, ``Scaling up
Trustless DNN Inference with Zero-Knowledge Proofs,'' arXiv:2210.08674,
2022.

{[}HCC{]} NVIDIA Corporation, ``Confidential Compute on NVIDIA Hopper
H100,'' Whitepaper WP-11459-001 v1.0, 25 July 2023.

{[}AR4SI{]} E. Voit, H. Birkholz, T. Hardjono, T. Fossati, V. Scarlata,
``Attestation Results for Secure Interactions,''
draft-ietf-rats-ar4si-10, IETF, work in progress, 18 May 2026
(Internet-Draft).

{[}EAR{]} T. Fossati, E. Voit, S. Trofimov, H. Birkholz, ``EAT
Attestation Results,'' draft-ietf-rats-ear-04, IETF, work in progress,
26 May 2026 (Internet-Draft, intended status: Standards Track).

{[}TCG-TPM2{]} Trusted Computing Group, ``Trusted Platform Module
Library Specification, Family 2.0'' (Parts 0--4).

{[}TCG-PCCFW{]} Trusted Computing Group, ``PC Client Platform Firmware
Profile Specification, Family 2.0,'' Version 1.06 Revision 52, 4
December 2023.

{[}swtpm{]} S. Berger et al., swtpm --- libtpms-based TPM 1.2/2.0
emulator. \url{https://github.com/stefanberger/swtpm} (run used version
0.7.3).

\end{document}